\documentclass[11pt,a4paper]{article}
\usepackage{jcappub}
\usepackage{color}
\usepackage{amsfonts, amssymb,amsmath,bm}
\usepackage{slashed}
\usepackage{braket}
\usepackage[normalem]{ulem}
\usepackage{comment}
\usepackage{xspace} 
\usepackage{siunitx}
\usepackage{multirow}
\usepackage{subcaption}
\usepackage{placeins}
\usepackage{afterpage}
\usepackage{booktabs}

\definecolor{ForestGreen}{rgb}{0.13, 0.55, 0.13}
\definecolor{airforceblue}{rgb}{0.36, 0.54, 0.66}
\definecolor{orange}{rgb}{1.0, 0.5, 0.0}
\definecolor{amethyst}{rgb}{0.6, 0.4, 0.8}
\definecolor{awesome}{rgb}{1.0, 0.13, 0.32}
\definecolor{chromeyellow}{rgb}{1.0, 0.65, 0.0}

\newcommand{\eos}{\textsc{Eos}\xspace}
\newcommand{\dataset}{\textsc{Dataset}\xspace}

\newcommand{\sympy}{\textsc{SymPy}\xspace}
\title{A Model for the Squeezed Bispectrum in the Non-Linear Regime}

\author[a,b,c]{Matteo Biagetti,}
\author[d]{Juan Calles,}
\author[e]{Lina Castiblanco,}
\author[d]{Katherine Gonz\'alez,}
\author[d]{and Jorge Nore\~na.}

\affiliation[a]{AREA Science Park, Padriciano 99, 34149, Trieste}
\affiliation[b]{Institute for Fundamental Physics of the Universe, Via Beirut 2, 34151 Trieste, Italy}
\affiliation[c]{SISSA - International School for Advanced Studies, Via Bonomea 265, 34136 Trieste,  Italy}
\affiliation[d]{Instituto de F\'isica, Pontificia Universidad Cat\'olica de Valpara\'iso, Casilla 4950, Valpara\'iso, Chile}
\affiliation[e]{School of Mathematics, Statistics and Physics, Newcastle University, Herschel Building, NE1 7RU Newcastle-upon-Tyne, U.K.}

\abstract{We present a model for the squeezed dark matter bispectrum, where the short modes are deep in the non-linear regime. We exploit the consistency relations for large-scale structures combined with a response function approach to write the squeezed bispectrum in terms of a few unknown functions of the short modes. We provide an ansatz for a fitting function for these response functions, checking that the resulting model is reliable when compared to the one-loop squeezed bispectrum. We then test the model against measured bispectra from numerical simulations for short modes ranging between $k \sim 0.1 \, h/$Mpc, and $k \sim 0.7 \, h/$Mpc at redshift $z=0$. To evaluate the goodness of the fit of our model we implement a non-Gaussian covariance and find agreement within one standard deviation of the simulated data.}

\begin{document}

\maketitle

\flushbottom 

\newpage

\section{Introduction}

Data from Large Scale Structure (LSS) surveys such as BOSS \cite{BOSS:2012dmf} and DESI \cite{DESI:2016fyo}, and the upcoming Euclid \cite{ Amendola:2016saw}, LSST \cite{Zhan:2017uwu}, SKA \cite{SKA:2018ckk}, SPHEREx \cite{Dore:2014cca} is and will be analyzed by comparing it with a model. Cosmological parameters and other physical information are extracted by looking for the values of the model parameters that best describe the data. It is therefore crucial to push the models to be reliable on the whole range of scales that are measured with high signal-to-noise by the surveys. Common approaches use perturbation theory to model the large-scale clustering using the redshift space galaxy power spectrum and bispectrum (see e.g. \cite{Bernardeau:2001qr,Baumann:2010tm,Carrasco:2012cv,Desjacques:2016bnm,Matsubara:2008wx,Porto:2013qua}). State-of-the-art techniques currently give a good description of BOSS data up to wave modes  $k\sim 0.2\,h/\text{Mpc}$ \cite{DAmico:2022osl} at low redshift. 

In this work, we want to push beyond this mildly non-linear scale, and model an observationally relevant quantity, the bispectrum, at scales smaller than what is accessible with perturbation theory. To do so, we exploit the non-perturbative character of the consistency relations of the large-scale structures \cite{Peloso:2013zw,Kehagias:2013yd}, and the response function approach \cite{Valageas:2013zda,Wagner:2014aka,Chiang:2014oga,Wagner:2015gva,Barreira:2017sqa}. These methods exploit the symmetries of the problem to greatly constrain the allowed functional form of the bispectrum in squeezed configurations, i.e. for configurations where one mode is much smaller than the other two.

Our motivation for studying the squeezed bispectrum is the search for primordial non-Gaussianity (PNG). One of the most expected measurements coming from the next generation of observations of the LSS is the correlation between a very long-wavelength perturbation of the primordial metric fluctuations with two small wave-length ones (the squeezed limit of the primordial bispectrum). If measured, it would rule out all single-field slow-roll inflationary models \cite{Maldacena:2002vr,Creminelli:2004yq,Creminelli:2011rh,Creminelli:2012ed}, since for single-field slow-roll models this correlation is trivial, given by a change of frame. The squeezed limit of the galaxy bispectrum is also completely fixed by a change of frame,  even if the small scales are very non-linear \cite{Peloso:2013zw,Kehagias:2013yd}. This is referred to as the consistency relation of the LSS. An observation of a deviation from the LSS consistency relation would indicate the presence of additional light fields during inflation \cite{Creminelli:2013mca}.


The late-time consistency relation suggests the absence of divergent poles in the squeezed galaxy bispectrum going as $q^{-2}$ and $q^{-1}$ \cite{Peloso:2013zw,Kehagias:2013yd,Creminelli:2013mca}, where $q$ is the mode going to zero. 
This property has been exploited in recent works \cite{Esposito:2019jkb,Goldstein:2022hgr} where the consistency relation is used to measure the amplitude of local primordial non-Gaussianity $f_{NL}$ from the squeezed limit of the matter bispectrum in simulations. 

The response approach is a powerful method to describe the non-linear scales where perturbation theory breaks down. It describes the effect of a long wavelength density or tidal field perturbation on the small-scale $n$-point correlation function of density perturbations. The construction is also based  on the approximate symmetries of the large-scale structure dynamical equations and, in fact, goes beyond the consistency relation \cite{Valageas:2013zda}. The response of the small-scale $n$-point correlation functions to a long wavelength perturbation is encoded in response functions. These are the coefficients in the expansion of the  $n$-point correlation functions with respect to a long-wavelength linear density perturbation \cite{Wagner:2015gva}. In the case of the power spectrum, the responses to a long-wavelength density perturbation provide a good description of the  squeezed limit of $n+1$ correlation functions \cite{Barreira:2017sqa}. In particular, the bispectrum in the squeezed limit is well described by the response of the power spectrum to a long wavelength perturbation. Reference \cite{Barreira:2017sqa} provides an analytical ``bias-like" expansion for the matter power spectrum response in terms of local operators. They use standard perturbation theory (SPT) to extrapolate the $n=1,2,3$ response coefficients  by matching to the tree-level bispectrum and trispectrum in the squeezed limit. Power spectrum responses have been accurately measured in separate-universe N-body simulations \cite{Li:2014sga,Wagner:2014aka,Chiang:2014oga,Wagner:2015gva,Voivodic:2020bec}.    


The response approach has also been applied to the power spectrum covariance \cite{Takada:2013wfa,Li:2014sga,Barreira:2017kxd,Li:2017qgh}, the bispectrum covariance \cite{Chan:2017fiv,Barreira:2019icq}, the supersample lensing covariance \cite{Barreira:2017fjz}, the integrated shear 3-point correlation function \cite{Halder:2022vft,Gong:2023nzy},to quantify the power spectrum overdensity response \cite{Racz:2021wrt} and in the presence of PNG \cite{Castorina:2020blr}, among other applications.

In this work, we compare the approach of using response functions and the consistency relation with simulations deep in the non-linear regime. We write an expansion for the small-scale density in the presence of a large-scale perturbation using the response function approach at the field level, Eq.~\eqref{eq:dm-density}. We then use a simple fitting function for the response coefficients and check that it provides a good description of the bispectrum as measured from the simulations. Our main results are summarized in Fig.~\ref{fig:bispectrum-from-sim}.

The paper is organized as follows. In Sec.\ \ref{sec:methodology} we outline our methodology to model the dark matter density at small scales with a response function modulated by a long-mode. In Sec.\ \ref{sec:results} we give the results of our analysis by comparing our best fit model to measurement on a dark matter N-body simulation. We conclude in Sec.\ \ref{sec:conclusions}.

\section{Response function expansion}\label{sec:methodology}

In this section, we describe how the dark matter density contrast at small (potentially non-perturbative) scales $\delta(\bm{k})$ responds to a change in the long-wavelength gravitational potential. At small enough scales, there is no way to compute the density field analytically. However, since the large scale density field is in the perturbative regime, the shape that the coupling between scales can take is constrained.

We use a response function approach similar to the one in \cite{Valageas:2013zda,Wagner:2014aka,Chiang:2014oga,Wagner:2015gva,Barreira:2017sqa} with a slight difference: We write the response of the small density contrast to a long-wavelength perturbation of the gravitational potential $\Phi(\bm{q})$ directly at the field level.\footnote{We use $\bm{q}$ for the long-wavelength mode throughout this work.} We find that doing it this way clarifies the role of the underlying symmetries, such as the implication of assuming an adiabatic evolution. In this response approach, the coupling between scales is written in terms of a few unknown functions, which we call response coefficients. This is reminiscent of using form factors to compute the untractable part of an amplitude involving hadrons in that the symmetries reduce the non-perturbative unknowns to a few free functions that can be measured.

The expansion is guided by the following symmetries:
\begin{itemize}
\item {\bf Rotational invariance.} Since the short-wavelength density contrast perturbation is a scalar under rotations, all vector indices should be contracted in the expansion.
\item {\bf The equivalence principle.} (Sometimes called Galilean invariance in this context.) For Gaussian initial conditions, we know that there is a physical coupling only to second and higher derivatives of the gravitational potential $\Phi(\bm{q})$. The coupling to the gravitational potential itself should be absent, while the coupling to the first derivative of the gravitational potential is fixed by the equivalence principle \cite{Peloso:2013zw,Kehagias:2013yd,Creminelli:2013nua,Creminelli:2013mca}. In particular, if the evolution of the large-scale perturbation is adiabatic throughout the history of the universe, this coupling can be written exactly \cite{Peloso:2013zw,Kehagias:2013yd,Creminelli:2013mca}.
\end{itemize}
Furthermore, the expansion is done in terms of the long-wavelength field and its derivatives:
\begin{itemize}
\item Since we take the large-scale mode to be in the linear regime, we keep terms only linear in the long-wavelength gravitational potential perturbation and its derivatives.
\item Take the long-wavelength Fourier mode to be $\bm{q}$, and the short wavelength modes to be of order $\sim k$. Derivatives are suppressed by $q/k$ or $q/k_{NL}$, where $k_{NL}$ is the non-linear scale. Since our focus will be on scales such that $q \ll k$ (the squeezed limit) and such that the long mode is linear (that is, $q \ll k_{NL}$), we keep only terms with the lowest order in derivatives. From the discussion above, we know that it should be the second derivative of the gravitational field $\partial_i\partial_j\Phi$. Through the Poisson equation, we can relate these to derivatives of the dark matter density contrast, apart from some constants $q^2 \Phi(q) \sim \delta(\bm{q})$, and $q^i q^j \Phi(\bm{q}) \sim \hat{q}^i \hat{q}^j \delta(\bm{q})$.
\end{itemize}

We, therefore, expand the short-wavelength density contrast in the presence of a long-wavelength perturbation as
\begin{multline}\label{eq:dm-density}
\delta(\bm{k})|_{\Phi_L} = \delta(\bm{k})|_{\Phi_L = 0} + \frac{(\bm{k} - \bm{q})\cdot\bm{q}}{q^2}\delta(\bm{q})\delta(\bm{k} - \bm{q}) \\ + \delta(\bm{q})\Delta_1(\bm{k}) +  \hat{q}^i \hat{q}^j \delta(\bm{q}) \hat{k}^i \hat{k}^j\Delta_\theta(\bm{k})  + \mathcal{O}(\delta(q)^2, (q/k)\delta(q))\,,
\end{multline}
where $\Delta_1(\bm{k})$ and $\Delta_\theta(\bm{k})$ are short-scale responses,\footnote{In this kind of expansions it is customary to use $\delta(\bm{q})$ and $(\hat{q}^i\hat{q}^j - \delta^{ij}/3)\delta(\bm{q})$ rather than the fields we use. This is to facilitate the physical interpretation of the different terms and to reduce degeneracies among them. We find it algebraically and numerically easier to work with our equivalent basis. Changing between the two bases is trivial.}
and $\delta(\bm{k})$ should be understood as the full non-linear density contrast. The second term in the expansion corresponds to the  consistency relation term \cite{Peloso:2013zw,Kehagias:2013yd}, which contains no dynamics and is fixed by the change from a local to a global coordinate frame \cite{Pajer:2013ana, Creminelli:2013mca}. The short-scale responses $\Delta_1(\bm{k})$, $\Delta_\theta(\bm{k})$ encode how the short scales respond to a long-wavelength mode. They are stochastic, in the sense that they depend on the initial conditions for the density contrast $\delta(\bm{k})$. However, there is no reason why they should be proportional to $\delta(\bm{k})$ at the non-linear level. As we show below, their form can be written explicitly at fixed order in perturbation theory. 
However, at scales where perturbation theory breaks down, they cannot be predicted a priori. 

Using the expansion of Eq.~\eqref{eq:dm-density} we can write an expression for the squeezed bispectrum
\begin{align}\label{eq:squez-bispectrum}
\lim_{q \rightarrow 0} \langle \delta(\bm{q}) \delta(\bm{k}_1) \delta(\bm{k}_2) \rangle' &= \lim_{q \rightarrow 0} \Big[\langle \delta(\bm{q}) \delta(\bm{k}_1) \delta(\bm{k}_2)|_{\Phi_L} \rangle' + \langle \delta(\bm{q}) \delta(\bm{k}_1)|_{\Phi_L} \delta(\bm{k}_2) \rangle' \Big ]\nonumber\\
&= P_m(q)P_m(k_1)\bigg[\frac{\bm{k}_1.\bm{q}}{q^2} + R_1(k_1) + R_\theta(k_1)(\hat{k}_1.\hat{q})^2\bigg] + (1 \leftrightarrow 2)\,,
\end{align}
where the prime indicates the momentum conservation delta has been removed and 
\begin{equation}
R_i(k) = \frac{\langle \delta(\bm{k}) \Delta_i(-\bm{k})\rangle'}{P_m(k)}\,,
\end{equation}
with $i=1,\theta$ denoting the different response coefficients. Corrections to Eq.~\eqref{eq:squez-bispectrum} come from terms of higher order in the ratio $q/k$ or $q/k_{NL}$, as well as terms which are not proportional to $P_m(q)$. In particular, we expect it to break down if the scales are such that $q$ is close to $k$ or $P_m(q)$ is smaller than $P_m(k)$.

\subsection{Analytic check with perturbation theory}
We can now check that the expansion makes sense by explicitly computing the squeezed bispectrum at a given perturbative order in standard perturbation theory (SPT), and matching the response coefficients.

\paragraph{Tree-level bispectrum.} To illustrate the method we start by checking the density contrast at second order, which is what we need in order to write the tree-level bispectrum. From standard perturbation theory, we get
\begin{equation}
\delta(\bm{k})|_{\Phi_L} = \delta_\ell(\bm{k}) + 2F_2(\bm{q}, \bm{k} - \bm{q})\delta_\ell(\bm{q})\delta_\ell(\bm{k} - \bm{q}) + \mathcal{O}(\delta_\ell^3, q/k\delta_\ell(q)\delta_\ell(k))\,,
\end{equation}
where the factor of 2 in the second term comes from the fact that either mode can be the long mode, the subscript $\ell$ denotes a quantity in the linear approximation, and the second-order kernel is given by
\begin{equation}
F_2(\bm{k}_1, \bm{k}_2) = \frac{5}{7} + \frac{\hat{k}_1\cdot\hat{k}_2}{2}\left(\frac{k_1}{k_2} + \frac{k_2}{k_1}\right) + \frac{2}{7}(\hat{k}_1\cdot\hat{k}_2)^2\,.
\end{equation}
Taking the $q \rightarrow 0$ limit we get
\begin{equation}
F_2(\bm{q}, \bm{k} - \bm{q}) = \frac{(\bm{k} - \bm{q})\cdot\bm{q}}{2q^2} + \frac{5}{7} + \frac{2}{7}(\hat{q}\cdot\hat{k})^2\,.
\end{equation}
This agrees with the expansion in Eq.~\eqref{eq:dm-density} by taking
\begin{equation}
\Delta_1(\bm{k}) = \frac{10}{7}\delta_\ell(\bm{k})\,,
\quad \Delta_\theta(\bm{k}) = \frac{4}{7}\delta_\ell(\bm{k})\,.
\end{equation}

The bispectrum should be well-described by the response approach in the limit in which $q \ll k$, and $P(q) \gg P(k)$. In that limit, we can write the tree level bispectrum as
\begin{align}
\lim_{q \rightarrow 0} \langle \delta(\bm{q}) \delta(\bm{k}_1) \delta(\bm{k}_2) \rangle' &\approx 2F_2(\bm{q}, \bm{k}_1)P_\ell(q)P_\ell(k_1) + 2F_2(\bm{q}, \bm{k}_2)P_\ell(q)P_\ell(k_2) \\
&= \left(\frac{\bm{k}_1.\bm{q}}{q^2} + \frac{10}{7} + \frac{4}{7}(\hat{q}.\hat{k}_1)^2\right)P_\ell(q)P_\ell(k_1) + (1 \leftrightarrow 2)\,.
\end{align}
This agrees with the expression in Eq.~\eqref{eq:squez-bispectrum} by taking
\begin{equation}
R_1(k) = \frac{10}{7}\,,\quad R_\theta(k) = \frac{4}{7}\,.
\end{equation}

\begin{figure}    
     \begin{subfigure}[b]{0.5\textwidth}
         \includegraphics[width=\textwidth]{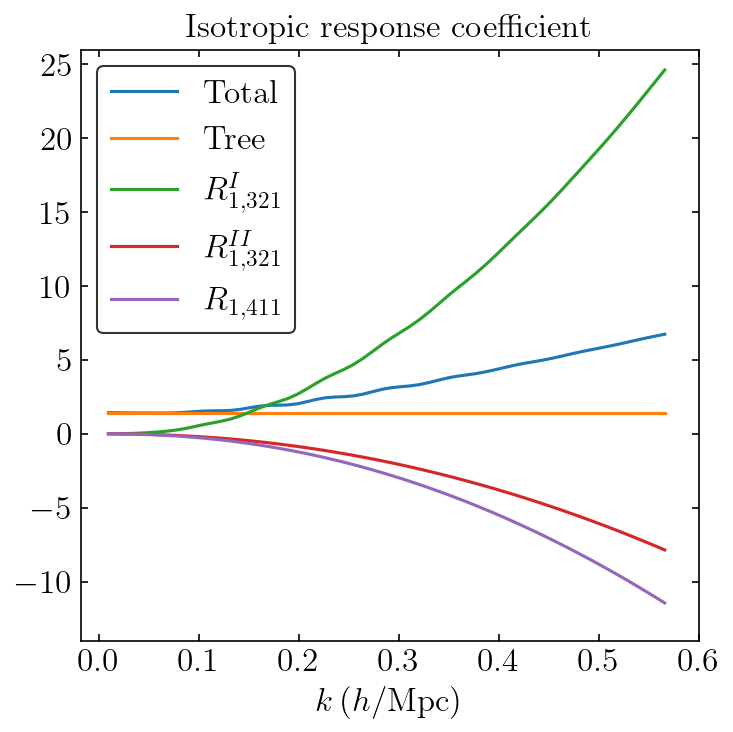}
     \end{subfigure}
     \begin{subfigure}[b]{0.5\textwidth}
         \includegraphics[width=\textwidth]{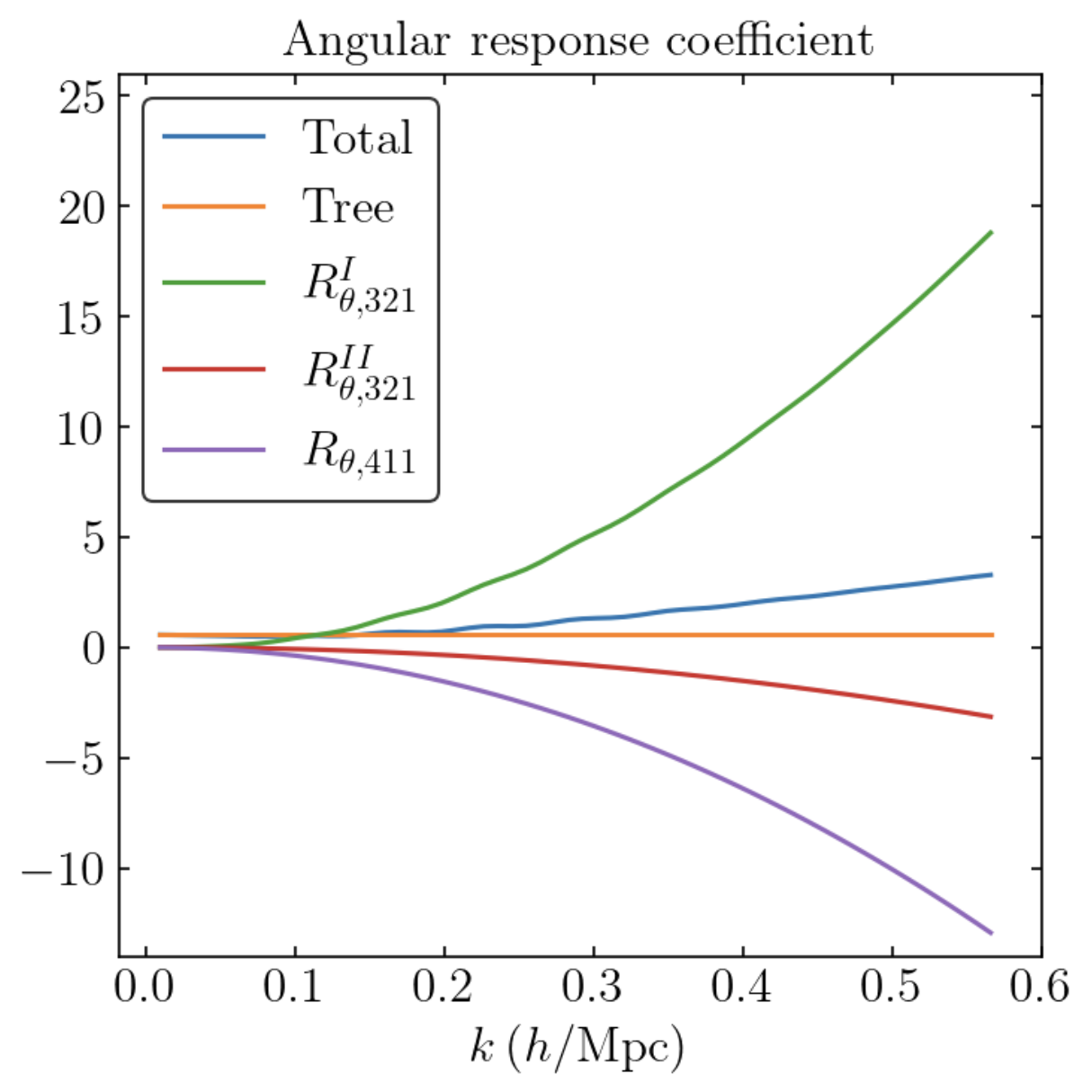}
     \end{subfigure}
        \caption{Isotropic (left) and angular (right) response coefficients computed from the full integration of each individual term in Eq.~\eqref{eq:1loop-squez-bis} of the tree level and one-loop squeezed bispectrum.}
        \label{fig:response-PT}
\end{figure}

\paragraph{One-loop bispectrum.} At tree-level, the response coefficients are trivially proportional to the non-linear density contrast. Since Eq.~\eqref{eq:squez-bispectrum} is expected to be valid non-perturbatively, it is interesting to repeat the exercise at the one-loop level. For the one-loop bispectrum, we need the density contrast up to the fourth order \footnote{Notation $$\int\,\frac{d^3k}{(2\pi)^3}= \int_{\bm{k}}$$}
\begin{align}
\delta(\bm{k})|_{\Phi_L} & = \delta_\ell(\bm{k}) + 2F_2(\bm{k} - \bm{q}, \bm{q}) \delta_\ell(\bm{k} - \bm{q})\delta_\ell(\bm{q}) \nonumber\\ 
&+ 3\int_{\bm{p_1},\bm{p_2}}  (2\pi)^3 \delta_D(\bm{k} - \bm{q} - \bm{p}_{12}) F_3(\bm{p}_1, \bm{p}_2, \bm{q})\delta_\ell(\bm{p}_1)\delta_\ell(\bm{p}_2)\delta_\ell(\bm{q}) \nonumber\\
&+ 4 \int_{\bm{p_1},\bm{p_2},\bm{p_3}}  (2\pi)^3 \delta_D(\bm{k} - \bm{q} - \bm{p}_{123}) F_4(\bm{p}_1, \bm{p}_2, \bm{p}_3, \bm{q})\delta_\ell(\bm{p}_1)\delta_\ell(\bm{p}_2)\delta_\ell(\bm{p}_3)\delta_\ell(\bm{q})
\end{align}
where the perturbation theory kernels $F_3$ and $F_4$ are computed similarly to $F_2$ and whose explicit expressions can be found in \cite{Goroff:1986ep,Bernardeau:2001qr}. The corresponding squeezed bispectrum at one loop
is
\begin{align}\label{eq:1loop-squez-bis}
\lim_{q \rightarrow 0} \langle \delta(\bm{q}) \delta(\bm{k}_1) \delta(\bm{k}_2) \rangle' &\approx 2F_2(\bm{q}, \bm{k}_1)P_\ell(q)P_\ell(k_1)\nonumber \\
&\phantom{=} + 6\int_{\bm{p}} F_2(\bm{p}, - \bm{p} - \bm{k}_2) F_3(-\bm{q}, \bm{p}, -\bm{p} - \bm{k}_2) P_\ell(p)P_\ell(|\bm{p} + \bm{k}_2|)P_\ell(q)\nonumber \\
&\phantom{=} + 6\int_{\bm{p}}  F_2(\bm{q}, \bm{k}_2) F_3(\bm{k}_2, \bm{p}, -\bm{p})P_\ell(k_2)P_\ell(p)P_\ell(q)\nonumber \\
&\phantom{=} + 12\int_{\bm{p}} F_4(\bm{q}, \bm{k}_2, \bm{p}, -\bm{p})P_\ell(k_2)P_\ell(q)P_\ell(p)\nonumber\\
&\phantom{=} + (1 \leftrightarrow 2)\,
\end{align}
where each of the terms is schematically defined as $B_{211}$, $B_{321}^I$, $B_{321}^{II}$ and $B_{411}$, respectively.\footnote{Note that we ignore the $B_{222}$ term of the usual one-loop bispectrum since we take the long mode to evaluated at linear scales. This term is suppressed by an additional power of the long-wavelength density contrast}. As above, we now want to find expressions for $R_1(k)$ and $R_\theta(k)$ that match  Eq.~\eqref{eq:1loop-squez-bis}. The general prescription to compute each contribution coming from Eq.~\eqref{eq:1loop-squez-bis} to the response coefficient $R_1(k)$ and $R_s(k)$ can be summarized as follows: 
\begin{itemize}
    \item Compute the squeezed limit bispectrum and write it as an expansion in $q$ for each kernel.
    \item The consistency relation is satisfied for each of the kernels. Subtract the consistency relation term at order  $q^0$ from the squeezed expression.
    \item Take the sub-leading terms to find the response coefficients. There are terms with no angular dependence, from which we can read $R_1(k)$. Other terms are proportional to $\mu^2 = (\hat{k}.\hat{q})^2$, from which we can read $R_\theta(k)$.
\end{itemize}
The explicit calculation is presented in Appendix~\ref{app:one-loop-response-coeff}. The resulting response coefficients are given in terms of loop integrals.\footnote{We compute these coefficients at one loop by performing loop integrations numerically using the Cuba library \cite{Hahn:2004fe}.} In Fig.~\ref{fig:response-PT} we show the ``isotropic'' response coefficient, $R_1$, and the ``angular'' response coefficient, $R_\theta$. From the expressions in that Appendix and the accompanying code, note that the response of the short scales to the long-wavelength potential is not proportional to the non-linear density contrast at small scales. In particular, the non-linear kernels involved are quite different from the kernels appearing in the loop integrals of the power spectrum. 

\subsection{Fitting function}
\label{sec:fitting-function}

While it is helpful to have analytic expressions for the response coefficients, they are valid only at the perturbative order they have been computed. We would like to have a prescription to fit data well beyond the  mildly non-linear regime. The strategy is to write down  fitting functions for the response coefficients and check their validity against the one-loop bispectrum, where we have the exact formulae. We then use these fits to compare with simulations in the next section. 

\begin{figure}
    \centering
    \includegraphics[width=0.485\textwidth]{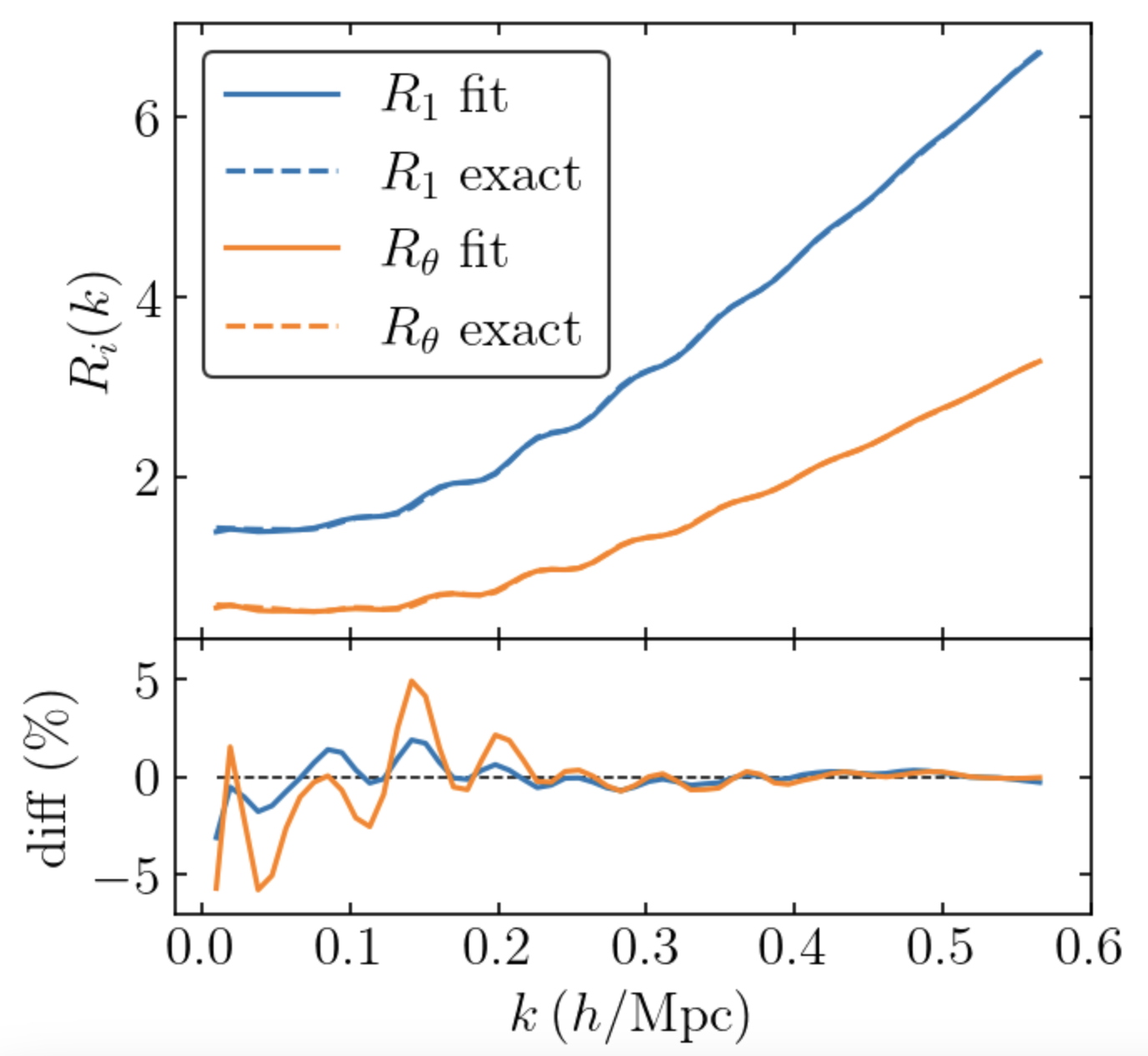}
    \includegraphics[width=0.47\textwidth]{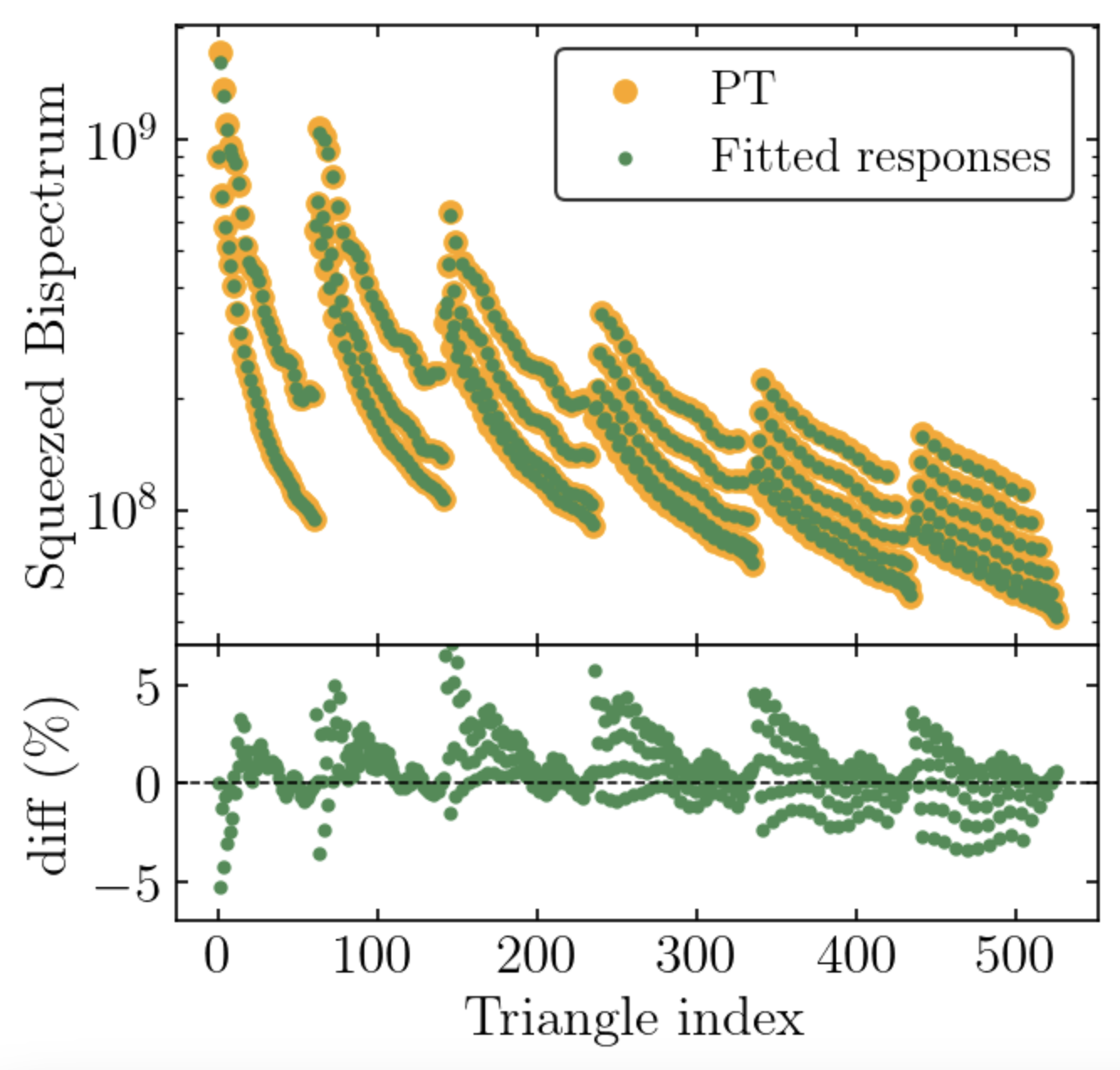}
    \caption{\emph{Left panel}: Fit to exact response coefficients using the fitting function of Eqs.~\eqref{eq:fitfun1} and \eqref{eq:fitfuntheta}. \emph{Right panel}: Comparison of the one-loop squeezed bispectrum calculated with the full perturbative formulae and using our model with the fitted response coefficients. Triangles considered have long modes in the range $q \sim 0.009$~to~$0.057\, h/$Mpc and short modes in the range $k \sim 0.094$~to~$0.565\, h/$Mpc. }
    \label{fig:respose-fit}
\end{figure}

From the results shown in Fig.~\ref{fig:response-PT} it is clear that the response functions are not simply proportional to the power spectrum of the short mode. Nevertheless, we still expect the response coefficients to be the result of two contributions: a smooth function of the short modes $k$ and a damped oscillatory term that is sourced by Baryon Acoustic Oscillations (BAO) with support in the range of scales from $k \approx 0.05 \,h/$Mpc to $k \approx
 0.5\, h/$Mpc. Consequently, we introduce the following fitting functions 
 \begin{align}
    \label{eq:fitfun1}
     R_1(k) &= \frac{10}{7} + S^1_1 k + S^1_2 k^2 + S^1_3 k^3 + (O^1_0 + O^1_1 k + O^1_2 k^2)P_{\rm nw}(k)e^{-\Sigma^2 k^2}\sin(\omega k + \phi)\,,\\
     \label{eq:fitfuntheta}
     R_s(k) &= \frac{4}{7} + S^\theta_1 k + S^\theta_2 k^2 + S^\theta_3 k^3 + (O^\theta_0 + O^\theta_1 k + O^\theta_2 k^2)P_{\rm nw}(k)e^{-\Sigma^2 k^2}\sin(\omega k + \phi)\,,
 \end{align}
 where $P_{\rm nw}(k)$ is the no-wiggle power spectrum, which we extract using the method of \cite{Vlah:2015zda} and $\Sigma$, $\omega$ and $\phi$ are the damping factor, the sound horizon scale and the phase of BAO oscillations, respectively. The first terms of Eqs.~\eqref{eq:fitfun1} and Eq.~\eqref{eq:fitfuntheta} are such that as $k \rightarrow 0$ we recover the tree level responses. We can now fit the free coefficients to the exact formulae at one-loop level. In Fig.~\ref{fig:respose-fit} we show the comparison of the fitted response coefficients with the ones computed analytically  (left) and the resulting fit to the squeezed one-loop bispectrum (right). Triangles considered to produce the one-loop squeezed bispectrum for the comparison have long modes in the range $q \sim 0.009$~to~$0.057\, h/$Mpc and short modes in the range $k \sim 0.094$~to~$0.565\, h/$Mpc. Note that the one-loop bispectrum at low redshift fails at much larger scales  than the scales considered here. Nevertheless, we chose a wide range that includes the full extent of BAO oscillations in order to verify that the fit works well. The assumption (checked against simulations in the following section) is that, while the free coefficients will change as we consider scales deep in the non-linear regime, the functional form of the fit can still describe the responses.   The fit turns out to be reliable within $5\%$ of the exact formulae. It is interesting to notice that we retrieve expected values for the BAO-related coefficients, $\Sigma \approx \mathcal{O}(5)$ Mpc$/h$ and $\omega \approx 100$ Mpc$/h$, which is expected given that the position of the BAO peak is protected from late-time non-linearities (see e.g. \cite{Baldauf:2015xfa,Blas:2016sfa}). Residual oscillations can be seen (lower panel of left Fig.~\ref{fig:respose-fit}), which hints that the fit is not properly modeling the oscillations. This could be easily improved, for instance modeling the $k$-dependence of the phase \cite{Baumann:2017gkg}, or including next-to-leading order corrections \cite{Blas:2016sfa}.  Nevertheless, for the range of scales involved in our simulation measurements the BAO are irrelevant, such that a better model does not affect the overall fit.

\section{Comparison with simulations deep in the non-linear regime}\label{sec:results}

We now show that our approach is valid in the deeply non-linear regime for the short scales. Indeed, the response expansion is based only on rotational invariance and the equivalence principle, which is valid at all scales. From the separate universe approach of \cite{Wagner:2014aka, Chiang:2014oga}, and from the one-loop calculation presented in the previous section, we expect the response coefficients to be smooth functions of the Fourier mode. We put this all together to fit the squeezed bispectrum in the deep non-linear regime using a few free parameters. We use the fitting function of Sec.~\ref{sec:fitting-function}, dropping the terms that describe the BAO. That is,
\begin{align}
    \label{eq:fitfun1noBAO}
     R_1(k) &= \frac{10}{7} + S^1_1 k + S^1_2 k^2 + S^1_3 k^3\,,\\
     \label{eq:fitfunthetanoBAO}
     R_s(k) &= \frac{4}{7} + S^\theta_1 k + S^\theta_2 k^2 + S^\theta_3 k^3\,.
 \end{align}
Since the long mode for our measured bispectra is below $0.05\,h/\text{Mpc}$, and for most of the triangles short mode is above $0.3\,h/\text{Mpc}$, the BAO are irrelevant. We checked that including BAO terms does not change the fit. However, they could be relevant again for dark matter tracers in redshift space.

\paragraph{General setup} For our check against simulated data, we use the \eos \dataset,\footnote{Full information about the \eos \dataset is found at \href{https://mbiagetti.gitlab.io/cosmos/nbody/eos/}{https://mbiagetti.gitlab.io/cosmos/nbody/eos/}.} a suite of full N-body simulations run with \textsc{Gadget}-2 \cite{Springel:2005mi}.  Initial conditions are implemented at $z_{in} = 99$ using second-order Lagrangian displacements with the \textsc{2LPTic} code \cite{Scoccimarro:1997gr,Crocce:2006ve}. We evolve $1536^3$ particles in a cubic periodic box of length $L=2$ Gpc/h up to redshift $z=0$. We consider matter snapshots from $11$ realizations for a total volume of $88 ({\rm Gpc}/h)^3$. The cosmology is flat $\Lambda$CDM with $\sigma_8 = 0.85$, $h = 0.7$ and $\Omega_m = 0.3$.

We are particularly interested in scales for which perturbation theory breaks down. We measure the matter power spectrum and bispectrum using a python version of the \textsc{PowerI4} code described in \cite{Sefusatti:2015aex}.\footnote{The PowerI4 code is found at \href{https://github.com/sefusatti/PowerI4}{https://github.com/sefusatti/PowerI4}.} We compute the bispectrum on binned triangles with modes that are multiples of a fundamental frequency  $k_f \sim 0.003\,h/$Mpc on bins of width $\Delta k_f = 3\, k_f$. The long modes range from $3\,k_f$ to $18\, k_f$ and the squeezing ratio is fixed to be larger than $10$ so that short modes range from $30\, k_f$ to $234 \,k_f$, corresponding to scales up to $k_{\rm max} = 0.735 \, h/$Mpc.

For the bispectrum model, we use Eq.~\eqref{eq:squez-bispectrum}, and for the response coefficients, we use the fitting functions of equations Eqs.~\eqref{eq:fitfun1noBAO} and \eqref{eq:fitfunthetanoBAO}. In this work, we use the linear power spectrum for the long mode, and the non-linear power spectrum measured from simulations for the short mode.\footnote{For parameter estimation, it would make sense to use the power spectrum measured from simulations for both. In that way, the non-Gaussian part of the covariance would be partially canceled, reducing the errors in the parameters.} We evaluate the model at the center of each bin. For this reason, we exclude triangles for which $q \geq |k_2 - k_1|$. In order to include them, one would have to average the model over each bin, since for these configurations only a few triangles inside the bin satisfy the triangle condition.

\paragraph{Fitting procedure} In order to fit, we look for the response coefficient that maximizes the likelihood 
\begin{equation}
    \text{ln}\mathcal{L} = -\frac{1}{2} (\bm{D}\cdot C^{-1}\cdot\bm{D})\,,
\end{equation}
where $\bm{D} = \bm{B}_{sim}-\bm{B}_{model}$ and for the covariance $C$ we use the theoretical covariance for squeezed triangles proposed in \cite{Biagetti:2021tua}
\begin{equation}
\label{eq:covariance}
C_{ij}^{B} \simeq \frac{\delta_{ij} }{k_f^3 N_{tr}^i}   P(q^i)P(k_1^i)P(k_2^i) +  \frac{k_f^2}{4\pi \Delta k k_1^i k_2^i} B(q^i,k_1^i, k_2^i)B(q^j, k_1^j, k_2^j) \,.
\end{equation}
Here, $i$, $j$ denote two different triangles, $P$ and $B$ are the power spectrum and bispectrum measured from simulations, $\Delta k$ is the size of the $k$ bin, and $N^i_{tr}$ is the number of independent triangles that contribute to the measurement of a given bispectrum configuration.\footnote{Note that in our model we used the measured power spectrum for the short modes. This means that we should not use the covariance of the bispectrum in the likelihood. Rather, we should use the covariance of $\bm{D}$. However, the difference between the two is of the same order as many terms neglected in Eq.~\eqref{eq:covariance}. If we used the measured power spectrum for the long mode, we would not be able to neglect the difference between the covariance of $\bm{D}$ and the bispectrum. Since doing that would partially cancel the covariance, this may be important if one wants to apply this approach to parameter estimation or model comparison.}

To the covariance, we add a theoretical error coming from the limitations of our approximation. Our ansatz for this error is
\begin{equation}
\label{eq:theory-error}
\sigma^2_{th} = \left(\frac{q}{k_1}\right) \hat{k}_1.\hat{q}\, P(q)P(k_1) + \left(\frac{q}{k_1}\right)^2 P(q)P(k_1) + (1 \leftrightarrow 2)\,.
\end{equation}
The first term in this expression is the expected order of magnitude of terms suppressed by $q/k$. Due to rotational invariance, we expect these terms to also be proportional to $\hat{q}.\hat{k}$. The second term is the expected order of magnitude of terms suppressed by $q^2/k^2$. In Appendix~\ref{app:theory-error}, we check this by comparing the one loop SPT bispectrum with the one obtained from the response model. The difference gives us an estimate of the size of terms neglected.

Since the parameters in the model appear linearly in the expression for the bispectrum, we analytically minimize the likelihood in order to find the best fit model.\footnote{In order to check that the BAO terms do not change the fit, we included  parameters which appear non-linearly, namely $\omega$, $\Sigma$, $\phi$. For given values of $\omega$, $\Sigma$, $\phi$, we can analytically minimize the likelihood. We therefore define a reduced likelihood function, depending on $\omega$, $\Sigma$, and $\phi$ which takes the minimum value of the likelihood function for fixed values of these three parameters. We then numerically minimize this reduced likelihood to obtain the best fit values of the parameters.}

\begin{figure}[ht]
    \centering
    \includegraphics[width=0.48\textwidth]{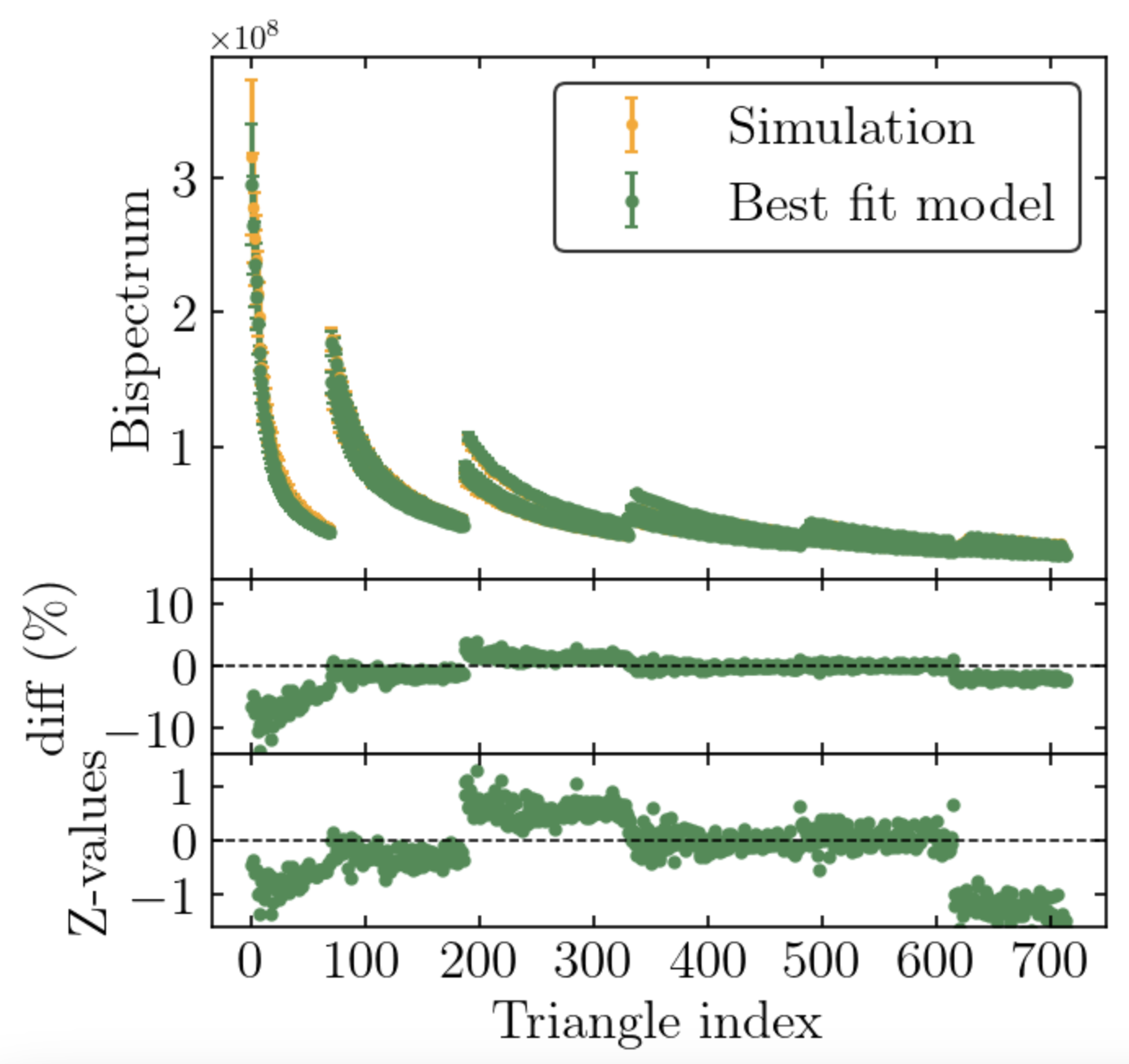}
    \includegraphics[width=0.48\textwidth]{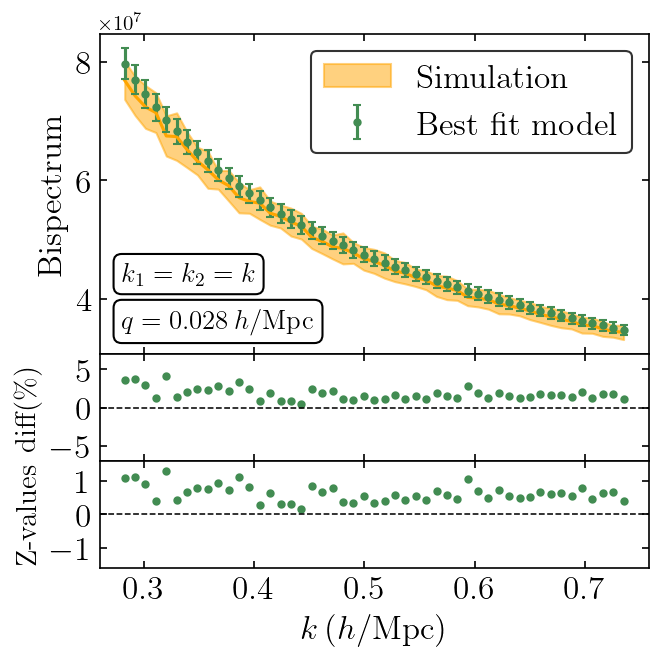}
    \caption{\emph{Left panel:} Comparison of our best fit model to simulation measurements of the squeezed bispectrum with long modes in the range $q \sim 0.009$ to $0.057\, h/$Mpc and short modes in the range $k \sim 0.094$ to $0.735\, h/$Mpc. Error bars for simulated data indicate the standard deviation computed over $11$ realizations, while for fitted data the error bars represent the variance from Eq.~\eqref{eq:covariance}. Lower panels show the percentage difference and the number of standard deviations between the simulation and best fit model Z-values $= (B_{model} - B_{sim})/\sigma$, respectively. \emph{Right panel:} Similar comparison for a subset of squeezed triangles for which $q = 0.028\, h/$Mpc as a function of $k_1 = k_2 = k$. }
    \label{fig:bispectrum-from-sim}
\end{figure}

\paragraph{Results} In Fig.~\ref{fig:bispectrum-from-sim} we plot the best fit response model along with the bispectrum measured from $11$ realizations. The yellow dots represent the average bispectrum among realizations. The yellow error bars represent the standard deviation estimated from the realizations. The green dots are the values of the best fit bispectrum modelled with response functions. The green bars are the standard deviation from a sum of Eqs.~\eqref{eq:covariance} and \eqref{eq:theory-error}. 
We see that the response model provides an excellent fit to the simulated bispectrum at deeply non-linear scales. The difference between the simulated points and the model is within roughly one standard deviation.

\begin{figure}[ht]
    \centering
    \includegraphics[width=0.48\textwidth]{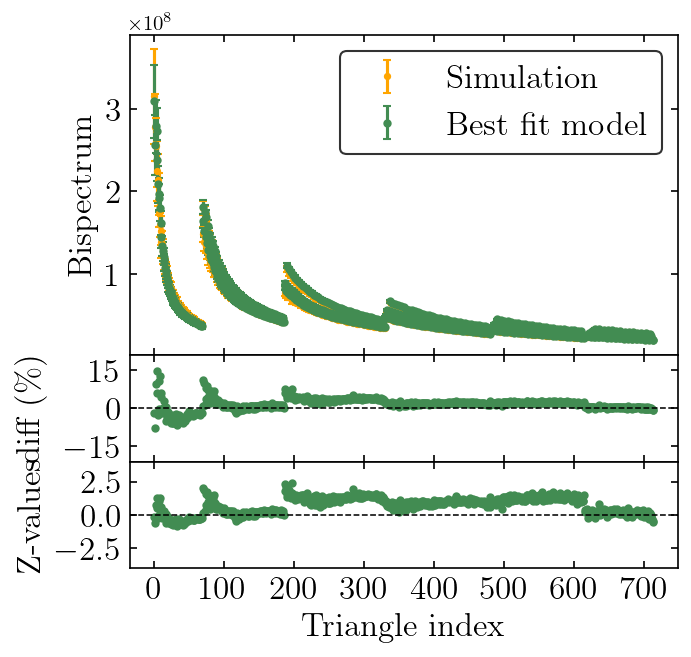}    \includegraphics[width=0.48\textwidth]
    {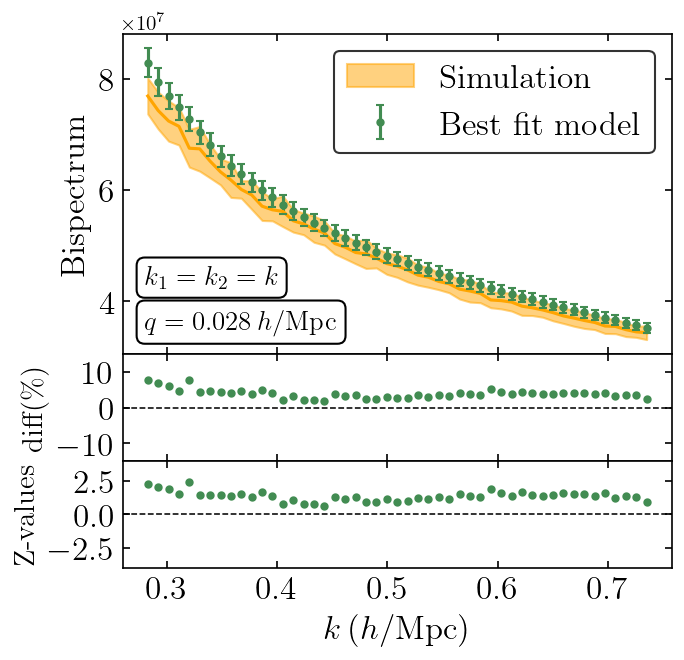}
    \caption{Same as Fig.~\ref{fig:bispectrum-from-sim}, except that the consistency relation term has been set to zero.}
    \label{fig:bispectrum-from-sim-noCR}
\end{figure}

Finally, it is interesting to check how much the consistency relation contributes to this model. In Fig.~\ref{fig:bispectrum-from-sim-noCR} we repeat the same plot as before, but we set the consistency relation term to zero. We fit the model from scratch in order to allow for the free parameters to try to mimic the consistency relation. We see that this model still provides a decent fit to the simulations, though many configurations systematically overestimate the bispectrum by $\sim 2\sigma$. 
The fact that the fit works is because at equal times the $1/q$ contribution of the consistency relation cancels. This can be seen by expanding the consistency relation term in the bispectrum
\begin{align}
\lim_{q \rightarrow 0} B(q, k_1, k_2) &\supset \frac{\bm{k}_1.\bm{q}}{q^2} P_m(q) P_m(k_1) + \frac{(-\bm{q} - \bm{k}_1).\bm{q}}{q^2}P_m(q) P_m(|\bm{q} + \bm{k}_1|) \nonumber \\
&\approx -2P_m(q)P_m(k_1) - (\hat{k}_1.\hat{q})^2 P_m(q)P'_m(k_1)\,.
\end{align}
Indeed, these terms are included as part of the response coefficients in \cite{Barreira:2017kxd}.





\section{Conclusions}\label{sec:conclusions}

We compared a model for the squeezed bispectrum in the deep non-linear regime with numerical large scale structure simulations, Fig.~\ref{fig:bispectrum-from-sim}. The model is based on the general response of a short-wavelength dark matter density contrast perturbation to a long-wavelength perturbation in the gravitational potential at the field level, Eq.~\eqref{eq:dm-density}. For this, we used the symmetries of the gravitational formation of structure, namely the equivalence principle \cite{Peloso:2013zw,Kehagias:2013yd,Creminelli:2013nua} and rotational invariance. Our approach is based on the approach of \cite{Barreira:2017sqa} with the slight difference that we write the expansion at the field level directly. The results obtained are valid even when the short-wavelength scale is very non-linear, where perturbation theory does not apply. This bears a similarity to the use of form factors in computing amplitudes involving hadrons: Symmetries allow one to write incomputable factors in terms of a few free functions. The final expression is given in terms of the response coefficients $R_1(k)$ and $R_\theta(k)$.

To describe these response coefficients, we use a polynomial function which also encodes the expected BAO oscillations, Eqs.~\eqref{eq:fitfun1} and \eqref{eq:fitfuntheta}. We then fit the free parameters in this function to a set of dark matter simulations. We use hundreds of squeezed configurations involving scales in the deep non-linear regime not accessible to perturbation theory. We see that our fit works well, with the simulation measurements being roughly within one $\sigma$ of the best fit model.

To be useful for observations, we need to write the response model in redshift space and for galaxies. We expect this to be relatively straightforward. Even taking shot noise into account, the evolution of the galaxy number density field should  satisfy the equivalence principle. Therefore, the response of this field to a long-wavelength perturbation of the gravitational field will still be given by an expression analogous to Eq.~\eqref{eq:dm-density}. We can then connect the long-wavelength gravitational field to a long-wavelength number density contrast perturbation using linear perturbation theory. We expect that in addition to our fitting parameters, there will be a few linear and quadratic bias coefficients.

This approach of describing the very squeezed bispectrum can be useful in constraining primordial non-Gaussianity. A violation of the consistency relation would be a smoking gun signal of the presence of additional fields during inflation \cite{Maldacena:2002vr,Creminelli:2004yq,Creminelli:2011rh,Creminelli:2012ed}, or other non-trivial physical processes. This has been exploited in \cite{Dalal:2007cu,Matarrese:2008nc} to argue that primordial non-Gaussianity can induce a divergence in the power spectrum, the so-called scale-dependent bias. For the bispectrum, it has been used in \cite{Esposito:2019jkb,Goldstein:2022hgr}, who look for the violation in an average over squeezed bispectra. It would be interesting to see how the approach presented here can be used in a similar manner. We can in principle use a larger set of squeezed triangles than \cite{Esposito:2019jkb,Goldstein:2022hgr} since we are not constrained by the averaging procedure, and we could easily extend our approach to galaxies in redshift space. Finally, it might be interesting to look for ways of extracting the coupling between short and long scales at the field level.

\section*{Acknowledgments}
 We thank Massimo Pietroni for useful discussions and Kevin Pardede for comments on a draft. We thank Joaquin Rohland for help in writing the code used for the analytical calculations. M.B. is supported by the Programma Nazionale della Ricerca (PNR) grant J95F21002830001 with the title "FAIR-by-design". J.C. is supported by ANID scholarship No. 21210008 and “Beca término de tesis PUCV, 2022.” J.N. is supported by FONDECYT Regular grant 1211545. L.C. is supported by the STFC Astronomy Theory Consolidated Grant ST/W001020/1 from UK Research $\&$ Innovation.

\appendix

\section{One-loop computation of the response coefficients}\label{app:one-loop-response-coeff}

We perform the explicit computation of the response coefficients using the one-loop SPT expansion. We make use of the \sympy python library  \url{https://www.sympy.org/en/features.html} \cite{10.7717/peerj-cs.103} to perform the algebra between the SPT kernels. We also used our own package for vector algebra, and for generating the SPT kernels, which is available at \url{https://github.com/jorgenorena/spt_kernels}. The explicit code to compute all the kernels of the response coefficients is also made available in that repository.

As a warm-up, let us verify that the one-loop SPT bispectrum satisfies the consistency relation. We know that we should obtain
\begin{equation}
\label{eq:CR}
\lim_{q \rightarrow 0} \langle \delta(\bm{q}) \delta(\bm{k}_1) \delta(\bm{k}_2) \rangle' = \frac{\bm{k}_1.\bm{q}}{q^2} P_\ell(q) P_m(k_1) + (1 \leftrightarrow 2) + \mathcal{O}\left(\frac{q^0}{k^0}\right)\,,
\end{equation}
where $P(k)$ is the non-linear power spectrum.

In order to satisfy the equivalence principle, the kernels should behave in the squeezed limit as
\begin{equation}
\lim_{q\rightarrow 0} F_2(\bm{q}, \bm{k}) = \frac{\bm{k}.\bm{q}}{2q^2} + \mathcal{O}(q^0)\,,
\end{equation}
\begin{equation}
\lim_{q \rightarrow 0}F_3(\bm{q}, \bm{p}_1, \bm{p}_2) = \frac{(\bm{p}_1.\bm{q} + \bm{p}_2.\bm{q})}{3q^2} F_2(\bm{p}_1, \bm{p}_2) + \mathcal{O}(q^0)\,,
\end{equation}
\begin{equation}
\lim_{q \rightarrow 0}F_4(\bm{q}, \bm{p}_1, \bm{p}_2, \bm{p}_3) = \frac{(\bm{p}_{123}.\bm{q})}{4q^2} F_3(\bm{p}_1, \bm{p}_2, \bm{p}_3) + \mathcal{O}(q^0)\,,
\end{equation}
where $\bm{p}_{123} \equiv \bm{p}_1 + \bm{p}_2 + \bm{p}_3$. From the explicit expressions of the kernels, one can verify that this is indeed the case.

Taking the squeezed limit of the one-loop bispectrum, Eq.~\eqref{eq:1loop-squez-bis}, using the expressions above, gives
\begin{align}
&\lim_{q \rightarrow 0} \langle \delta(\bm{q}) \delta(\bm{k}_1) \delta(\bm{k}_2) \rangle' \approx \frac{\bm{q}.\bm{k}_1}{q^2}P_\ell(q)P_\ell(k_1) \\
&\phantom{=} - 2\int_{\bm{p}_1,\bm{p}_2}(2\pi)^3\delta_D(\bm{k}_1 + \bm{q} - \bm{p}_1 - \bm{p}_2) \frac{(\bm{p}_1.\bm{q} + \bm{p}_2.\bm{q})}{q^2} (F_2(\bm{p}_1, \bm{p}_2))^2 P_\ell(p_1)P_\ell(p_2)P_\ell(q) \\
&\phantom{=} + 3\int_{\bm{p}}  \frac{\bm{q}.\bm{k}_2}{q^2} F_3(\bm{k}_2, \bm{p}, -\bm{p})P_\ell(k_2)P_\ell(p)P_\ell(q) \\
&\phantom{=} + 3\int_{\bm{p}}  \frac{\bm{q}.\bm{k}_2}{q^2} F_3(\bm{k}_2, \bm{p}, -\bm{p})P_\ell(k_2)P_\ell(q)P_\ell(p) \\
&\phantom{=} + (1 \leftrightarrow 2)\,.
\end{align}
After simplifying a bit, we see that we recover the terms appearing in the one-loop expression for the power spectrum
\begin{align}
P(k) &=  P_\ell(k) \nonumber \\
&\phantom{=}+ 2\int_{\bm{p}} \left(F_2(\bm{k} - \bm{p}, \bm{p})\right)^2 P_\ell(|\bm{k} - \bm{p}|)P_\ell(p) \nonumber \\
&\phantom{=}+ 6\int_{\bm{p}}  F_3(\bm{k}, \bm{p}, \bm{-p}) P_\ell(p) P_\ell(k)\,,
\end{align}
such that the consistency relation, Eq.~\eqref{eq:CR}, is satisfied. The second term in this expression is customarily called $P_{22}$, and the third term is called $P_{13}$.

We now go one order higher in $q/k$. At this order, Eq.~\eqref{eq:1loop-squez-bis} is not enough to describe the bispectrum. We obtain the contribution from the response coefficients, which we can extract from the resulting expressions. We look at each term in turn.

\subsection*{$\bm{B_{321}^{II}}$}

We begin by studying the simplest contribution to the one-loop bispectrum, from Eq.~\eqref{eq:1loop-squez-bis}
\begin{align}
\lim_{q \rightarrow 0} B_{321}^{II}(q, k_1, k_2) &= \lim_{q \rightarrow 0} 6\int_{\bm{p}}  F_2(\bm{q}, \bm{k}_2) F_3(\bm{k}_2, \bm{p}, -\bm{p})P_\ell(k_2)P_\ell(p)P_\ell(q)  + (1 \leftrightarrow 2)\nonumber \\
&= 6 \left(\lim_{q \rightarrow 0} F_2(\bm{q}, \bm{k}_2)\right) P_\ell(k_2)P_\ell(q)\int_{\bm{p}}  F_3(\bm{k}_2, \bm{p}, -\bm{p})P_\ell(p)   + (1 \leftrightarrow 2) \nonumber \\
&= \left(\lim_{q \rightarrow 0} F_2(\bm{q}, \bm{k}_2)\right) P_\ell(k_2)P_\ell(q) P_{13}(k_2)   + (1 \leftrightarrow 2)\,.
\end{align}
We thus obtain finally
\begin{equation}
\lim_{q \rightarrow 0} B_{321}^{II}(q, k_1, k_2) = \left(\frac{\bm{k}_2.\bm{q}}{q^2} + \frac{5}{7} + \frac{2}{7}(\hat{k}_2.\hat{q})^2\right) P_\ell(k_2)P_\ell(q) P_{13}(k_2)   + (1 \leftrightarrow 2)\,.
\end{equation}
From this we can read the contribution to the response coefficients
\begin{equation}
R_{1,321}^{II} = \frac{5}{7}\frac{P_{13}(k)}{P_\ell(k)}\,,\quad R_{\theta,321}^{II} = \frac{2}{7}\frac{P_{13}(k)}{P_\ell(k)}\,.
\end{equation}
These are the expressions plotted in Fig.~\ref{fig:response-PT}.

\subsection*{$\bm{B_{321}^{I}}$}

Other contributions to the response coefficients can't be so neatly written. Let us consider the contribution from $B_{321}^{I}$, from Eq.~\eqref{eq:1loop-squez-bis}
\begin{multline}
B_{321}^{I}(q, k_1, k_2) = 6 P_\ell(q)\int_{\bm{p}} F_2(\bm{p}, -\bm{k}_2 - \bm{p}) F_3(-\bm{q}, \bm{p}, -\bm{k}_2 - \bm{p}) P_\ell(p)P_\ell(|\bm{k}_2 + \bm{p}|) \\ + (1 \leftrightarrow 2)\,.
\label{eq:B321I}
\end{multline}
In order to extract the response coefficients, we first subtract the contribution of this loop to the consistency relation. That is, we're interested in computing the leading term in
\begin{equation}
\lim_{q \rightarrow 0}\left[B_{321}^{I}(q, k_1, k_2) - \left(\frac{\bm{k}_2.\bm{q}}{q^2}P_{22}(k_2)P_\ell(q) + (1\leftrightarrow 2)\right)\right]\,.
\end{equation}
The kernels in the integrand in Eq.~\eqref{eq:B321I} depend on the magnitudes of $p$, $q$, and $k_2$, along with the dot products $\mu_q \equiv \hat{q}.\hat{p}$, $\mu_k \equiv \hat{p}.\hat{k}_2$, and $\mu \equiv \hat{k}_2.\hat{q}$. 

In order to evaluate the integral, we chose 
$$
\bm{k}_2 = k_2(0, 0, 1),\ \bm{q} = q(\sqrt{1 - \mu^2}, 0, \mu),\ \bm{p} = p(\sqrt{1 - \mu_k^2}\cos\phi, \sqrt{1 - \mu_k^2}\sin\phi, \mu_k)\,.
$$
From these, we get
$$
\mu_q = \sqrt{1 - \mu^2}\sqrt{1 - \mu_k^2}\cos\phi + \mu_k\mu\,.
$$
We wish to integrate over $p$, $\mu_k$, and $\phi$. Since the arguments of the power spectra do not depend on $\phi$, the integral over this angle can be done analytically. Since this is rather cumbersome, we used our symbolic computation code to perform the integral. 

We checked that the resulting expression only contains terms which are independent of $\mu$ or terms which are quadratic in $\mu$. The former give us an expression for $R_{1,321}^{I}$, and the latter an expression for $R_{\theta,321}^{I}$. These are still in terms of integrals over $p$ and $\mu_k$, which we perform numerically using the Cuba library \cite{Hahn:2004fe}. We plot the results in Fig.~\ref{fig:response-PT}.

\subsection*{$\bm{B_{411}}$}

Finally, let us consider the contribution from $B_{411}$. From Eq.~\eqref{eq:1loop-squez-bis}
\begin{equation}
B_{411}(q, k_1, k_2) = 4 \int_{\bm{p_1},\bm{p_2},\bm{p_3}}  (2\pi)^3 \delta_D(\bm{k} - \bm{q} - \bm{p}_{123}) F_4(\bm{p}_1, \bm{p}_2, \bm{p}_3, \bm{q})\delta_\ell(\bm{p}_1)\delta_\ell(\bm{p}_2)\delta_\ell(\bm{p}_3)\delta_\ell(\bm{q})\,.
\end{equation}
Once more, in order to extract the response coefficients, we first subtract the contribution of this loop to the consistency relation. That is, we're interested in computing the leading term in
\begin{equation}
\lim_{q \rightarrow 0}\left[B_{411}(q, k_1, k_2) - \frac{1}{2}\left(\frac{\bm{k}_2.\bm{q}}{q^2}P_{13}(k_2)P_\ell(q) + (1\leftrightarrow 2)\right)\right]\,.
\end{equation}
For the integrals are performed in the same way as for $B_{321}^{I}$. Once more, we obtain a piece which is independent of $\mu$ and a piece which is quadratic in $\mu$. The former give us an expression for $R_{1,411}$, and the latter an expression for $R_{\theta,411}$. We plot the results in Fig.~\ref{fig:response-PT}.

The fact that these rather involved kernels only give terms which are independent of $\mu$ or quadratic in $\mu$ validates the hypothesis behind the response function approach at one-loop order.

\section{Check of theoretical errors}
\label{app:theory-error}

\begin{figure}[h!]
    \centering
    \includegraphics[width=0.95\textwidth]{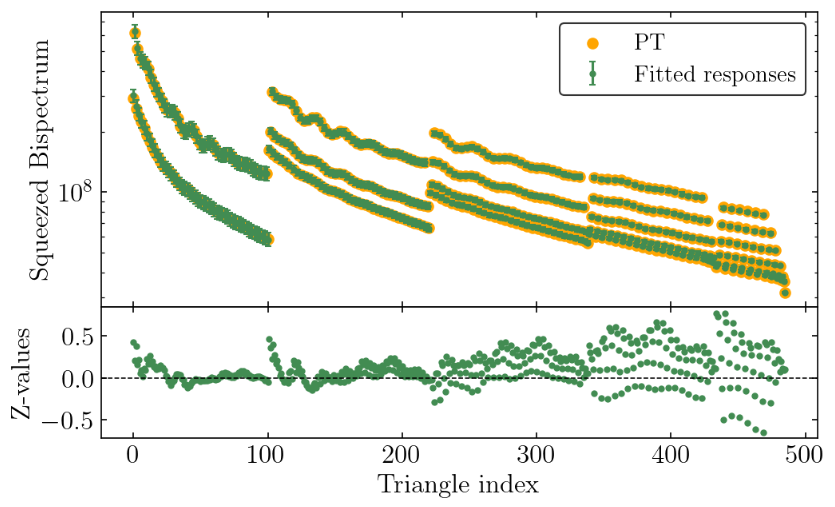}
    \caption{The one-loop squeezed bispectrum from the response model compared to the analytic solution, where we assign a theoretical error as specified in Eq. \eqref{eq:theory-error}. The triangle setup is the same as Fig. \ref{fig:respose-fit}. The lower panel indicates the standard deviation of the fit from the one-loop squeezed bispectrum. Z-values $= (B_{fit} - B_{PT})/\sigma_{th}$.}
    \label{fig:bispectrum-analytic-res}
\end{figure}

In this Appendix we check the expression for the theoretical errors in Eq.~\eqref{eq:theory-error}. For this, we compare in Fig.~\ref{fig:bispectrum-analytic-res} the full one-loop SPT bispectrum with the one obtained from the response model. In the lower panel, we show the Z-values, namely the difference between models divided by the theoretical error. We see that deviations are of the order of one $\sigma_{th}$. As such, our theoretical error robustly describes the terms in the one-loop expression which were ignored in the response model.

\bibliographystyle{JHEP}
\bibliography{references}
\end{document}